# U-Net Based Architecture for an Improved Multiresolution Segmentation in Medical Images

**Simindokht Jahangard**[a], **Mohammad Hossein Zangooei**[b], **Maysam Shahedi**[b,*]

[a] Amirkabir University of Technology, Department of Robotics Engineering, Department, Hafez Ave, Tehran, Iran, 1591634311

[b] The University of Texas at Dallas, Department of Bioengineering, 800 W. Campbell Road, Richardson, Texas, USA, 75080

**Abstract**.

**Purpose**: Manual medical image segmentation is an exhausting and time-consuming task along with high inter-observer variability. In this study, our objective is to improve the multi-resolution image segmentation performance of U-Net architecture.

**Approach**: We have proposed a fully convolutional neural network for image segmentation in a multi-resolution framework. We used U-Net as the base architecture and modified that to improve its image segmentation performance. In the proposed architecture (mrU-Net), the input image and its down-sampled versions were used as the network inputs. We added more convolution layers to extract features directly from the down-sampled images. We trained and tested the network on four different medical datasets, including skin lesion photos, lung computed tomography (CT) images (LUNA dataset), retina images (DRIVE dataset), and prostate magnetic resonance (MR) images (PROMISE12 dataset). We compared the performance of mrU-Net to U-Net under similar training and testing conditions.

**Results**: Comparing the results to manual segmentation labels, mrU-Net achieved average Dice similarity coefficients of 70.6%, 97.9%, 73.6%, and 77.9% for the skin lesion, LUNA, DRIVE, and PROMISE12 segmentation, respectively. For the skin lesion, LUNA, and DRIVE datasets, mrU-Net outperformed U-Net with significantly higher accuracy and for the PROMISE12 dataset, both networks achieved similar accuracy. Furthermore, using mrU-Net led to a faster training rate on LUNA and DRIVE datasets when compared to U-Net.

**Conclusions**: The striking feature of the proposed architecture is its higher capability in extracting image-derived features compared to U-Net. mrU-Net illustrated a faster training rate and slightly more accurate image segmentation compared to U-Net.

*****Maysam Shahedi**, E-mail: mshahedi@utdallas.edu

## 1 Introduction

Image segmentation plays a vital role in radiology and radiation oncology. An accurate and repeatable segmentation could be helpful to either better diagnosis of the abnormalities or more effective treatment of the diseases. For most of the current clinical procedures, medical image segmentation is done manually by an expert. However, manual image segmentation is usually a time-consuming and complicated process with high intra- and inter-observer variability [1]. A fast and reliable computer-assisted segmentation algorithm could speed-up the clinical procedures, decrease the costs, and increase the repeatability of the task. It could also facilitate image-guided procedures and make them more efficient by enabling fast intraoperative segmentation.

Many different medical image segmentation algorithms have been proposed in the literature that offer faster performance with higher repeatability than manual segmentation [2-6]. Some of these algorithms were developed based on handcrafted image feature extraction [7-9]. They were usually optimized for image segmentation on a specific image dataset. They demonstrated high segmentation accuracy, but their performance strongly depended on the selected feature set. Therefore, in order to adopt these algorithms for segmenting a new dataset, a new set of features must be selected, and the algorithm needs to be reoptimized for the new application. Besides, depending on the dimension of the feature set, feature extraction in these algorithms could be computationally intensive and the discriminative power of the features is questionable. Recently, deep learning algorithms have shown fast, reliable, and robust performance in various signal processing fields like medical image processing. Fully convolutional neural networks (FCNNs) have been widely used for image segmentation [10-12]. U-Net is an FCNN presented by Ronneberger et al. in 2015 [13] that has been widely used for medical image segmentation [14-17]. Yu et al. [18] proposed a new 3D CNN named DenseVoxNet for the cardiac and vascular MR image segmentation problem. To address volumetric brain segmentation, Chen et al. [19] presented a voxel residual network referred to as VoxResNet. They integrated a low-level image appearance feature, high-level context, and shape information to improve their performance. Yan et al. [20] proposed a model in which multi-level features are extracted and combined. Their network is designed to segment prostates on magnetic resonance (MR) images. Milletari et al. presented V-Net in 2016 [10] as another FCNN similar to U-Net used for 3D image volumes, which is also widely used in medical image segmentation. In both U-Net and V-Net, a multi-resolution paradigm for extracting the features was used. This helped the networks to extract features from different image scales. However, in both architectures, the features extracted in lower resolutions were not directly extracted from the input image. It makes it challenging for the network to extract low-frequency features that can be extracted easier from low-resolution versions of the image. Zeng et al. proposed a modified version of U-Net called RIC_Unet (residual-inception-channel attention-U-Net) for nuclei segmentation in histology images [21]. They employed a combination of three techniques to improve the segmentation performance; using residual blocks, extracting multi-scale features, and applying channel attention mechanism. Their network showed improved performance over the original U-Net. In another study, Jin et al. [22] presented a new architecture called RA-UNet and tested that for three-dimensional (3D) liver and tumor segmentation in computed tomography (CT) images and 3D brain tumor segmentation in magnetic resonance imaging (MRI). They used a hybrid approach to improve the U-Net architecture by using residual learning combined with attention mechanisms. Their network outperformed the prior algorithms presented for liver CT segmentation and showed comparable performance to prior algorithms had been presented for brain tumor segmentation in MRI. Their network outperformed the prior algorithms presented for liver CT segmentation and showed comparable performance to prior algorithms had been presented for brain tumor segmentation in MRI.

In this paper, we hypothesize that incorporating features extracted directly from down-sampled versions of the input image improves the segmentation performance of the U-Net architecture. To test our hypothesis, we used a cascading architecture to incorporate features directly extracted from lower-resolution versions of the input image into the network. Then we trained U-Net and the proposed network, hereafter called multiresolution U-Net (mrU-Net) using different datasets under the same training conditions. We tested the trained networks on the same test datasets and compared the performance of mrU-Net to that of U-Net.

The outline of this study is as follows. In Sec. 2, we briefly review the details of the datasets used in this study. We also present the architecture of mrU-Net and the error metrics used for performance evaluation. In Sec. 3, we provide explanations of the training and validation process for each dataset in detail and the achieved validation and test results for both mrU-Net and U-Net. We discuss and analyze the obtained outcomes in Sec. 4 and describe future directions and end with conclusions.

## 2 Materials and Methods

*2.1 Materials*

To evaluate our proposed method, four datasets were used in this study; the dataset of ISIC 2018 Skin Lesion Analysis Towards Melanoma Detection grand challenge [23, 24], lung computed tomography (CT) images (LUNA dataset) [25], the Digital Retinal Images for Vessel Extraction (DRIVE) dataset [26], and a prostate magnetic resonance imaging dataset (PROMISE12) provided by the MICCAI 2012 grand challenge [27]. For each of the images, one manual segmentation label was provided by an expert. Fig. 1 depicts some sample images and their segmentation labels from the datasets. For the skin, LUNA, and PROMISE12 datasets, we divided each dataset into two parts; 75% of the images were used for training purposes and the remaining 25% were used for final testing. About 10% of the training images were allocated as a validation dataset for evaluation of our model fit while tuning its hyperparameters. DRIVE dataset consists of 20 training and 20 test images. We selected four of the training images for validation. Due to limited data, we used data augmentation for the training and validation sets using all the eight unique combinations of 90, 180, and 270 degrees image rotation with vertical and horizontal image reflection to generated new training images. PROMISE12 dataset contains 3D images. In this study, we picked the 2D axial slices on which prostate tissue is available. Table 1 shows the number of total training, validation, and test images in each dataset, separately. For simplicity and consistency, all the images were resized to 512×512 pixels using bicubic interpolation and the original pixel intensity values were converted to eight-bit values.

**Table 1** The number of training, validation, and testing images for each dataset.

| Dataset | Training | Validation | Test | Bit Numbers | Initial image size | Final image size |
|---|---|---|---|---|---|---|
| **Skin lesion** | 1751 | 195 | 648 | 8-bit, RGB | Varies from 1024×768×3 to 512×384×3 | 512×512×3 |
| **LUNA** | 177 | 20 | 66 | 8-bit, grayscale | 512×512 | 512×512 |
| **DRIVE** | 16×8* | 4×8* | 20 | 8-bit, RGB | 512×512×3 | 512×512×3 |
| **PROMISE 12** | 526 | 58 | 185 | 8-bit, grayscale | 256×256, 320×320, 512×512 | 512×512 |

* Augmented data

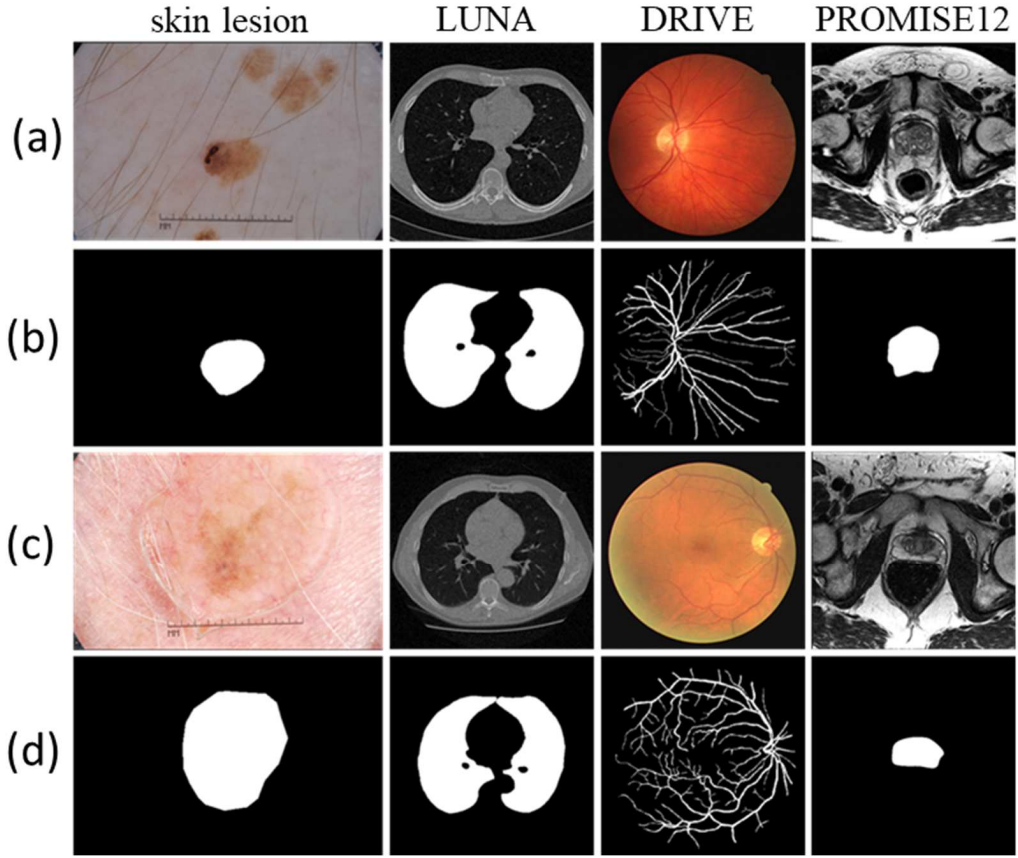

**Fig. 1** Sample images and their manual segmentation labels from the four datasets. Rows (a) and (c) show the images and rows (b) and (d) impart the ground truth of the corresponding images from the rows above them. Each column shows one dataset; form left to right, skin lesion, LUNA, DRIVE, and PROMISE12 dataset.

For each image, we normalized the intensity values to the range of zero to one. More specifically, for PROMISE12 images, we normalized the intensity of the pixels to the range of zero to one $\hat{I}(x,y)$ by using Eq. 1 in which $I(x,y)$ denotes the intensity value of the pixel, and $I_{min}$ and $I_{max}$ shows the minimum and maximum value of the pixels in the prostate region across all the training images, respectively. The same method was followed to obtain LUNA CT normalized images. For skin lesion and DRIVE datasets, pixel values in all the RGB channels were normalized to the range of zero to one by dividing the values by 255.

$$\hat{I}(x,y) = \frac{I(x,y) - I_{min}}{I_{max} - I_{min}} \qquad (1)$$

*2.2 Multiresolution U-Net (mrU-Net)*

*2.2.1 Network architecture*

In this study, a four-level U-Net was used as the basic end-to-end network architecture. We modified the architecture to improve its multiscale image feature extraction. We added some convolutional layers to each level of the contracting path (left side of the network) to extract some of the features directly from the down-sampled versions of the input image. Fig. 2 illustrates a general block diagram of the proposed network. We used zero-padding before each convolution layer to keep the size of the output of the layers the same as its input. We used 3×3 convolutional kernels for all the convolutional layers except the last layer of the network for which we used 1×1 convolutional kernel size. In the contracting path, we applied a cascading image down-sampling to the input image and after each down-sampling, the down-sampled image went through two convolutional layers. Then the feature maps were incorporated by concatenating them with the corresponding feature maps of the contracting path of the original U-Net network (see Fig. 2).

*2.2.2 Training*

To train the network, we used Adadelta [28] gradient-based optimizer and a loss function defined based on soft Dice Similarity Coefficient (sDSC) [10] defined as follows:

$$\text{Loss} = 1 - \text{sDSC} \qquad (2)$$

where,

$$sDSC = \frac{2\sum_i(p(I_i).G_i)}{\sum_i(p(I_i))+\sum_i(G_i)}, \tag{3}$$

$I_i$ and $G_i$ are the ith pixels of the input image and the reference binary label, respectively, and $p(I_i)$ is the output probability value corresponds to $I_i$. For both U-Net and mrU-Net, softmax output activation function is used which produces a probability value in the range (0, 1) for each of the pixels. The batch size and initial learning rate used for training the networks on each dataset is 16 and 1.0, respectively.

We applied thresholding with the threshold level of 50% to the output probability maps to build binary segmentation labels and compare them to the reference segmentation binary mask for evaluation purposes.

*2.3 Evaluation*

To evaluate the segmentation performance of the FCNNs, we used different error metrics such as Dice Similarity Coefficient (DSC) [29], sensitivity rate, and specificity rate [30] to compare the algorithm segmentation results against manual segmentation labels. Sensitivity rate and specificity rate were defined as follows:

$$Sensivity = \frac{TP}{TP+FN} \tag{4}$$

$$Specificity = \frac{TN}{TN+FP} \tag{5}$$

where TP (true positive) denotes the pixels correctly segmented as foreground, FP denotes the pixels incorrectly segmented as foreground, TN denotes pixels correctly segmented as background, and FN denotes the pixels incorrectly segmented as background. In this paper, we reported all the metrics in percentage.

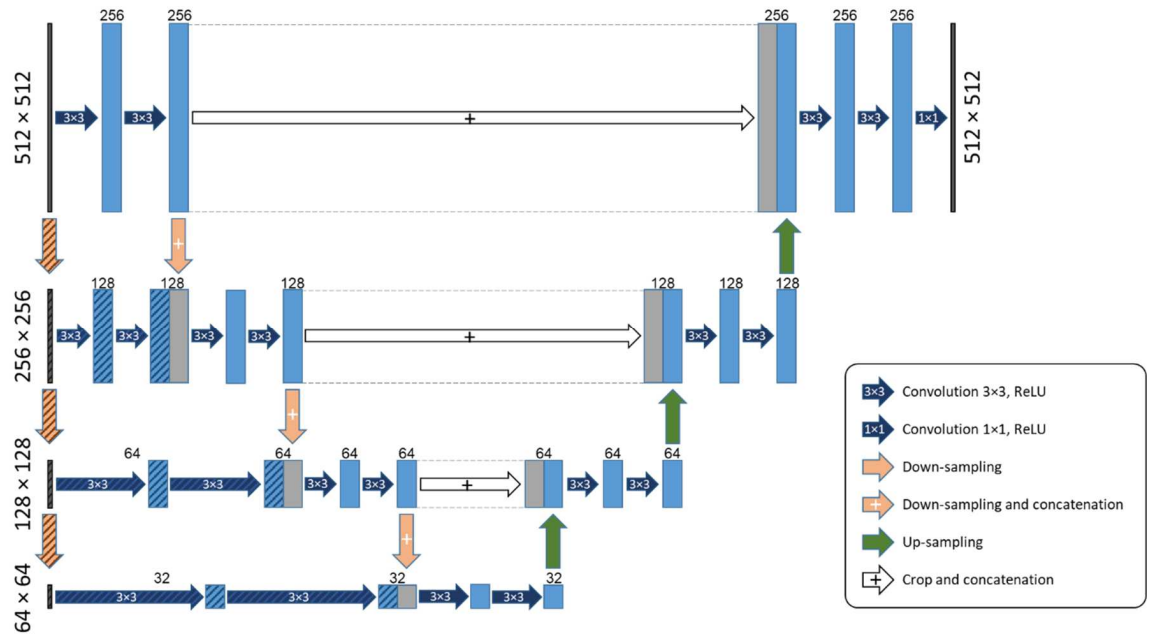

**Fig. 2** Four-level mrU-Net architecture. The hashed layers are added layers to the U-Net architecture. The numbers above the feature maps indicate the number of feature channels. The size of the original input image is 512×512 and it has either one channel for grayscale images or three channels for RGB images.

## 3   Experiments and Analytic Results

### 3.1  Training

For each network and each dataset, we continued training for 5000 epochs and chose the model that achieved the highest validation accuracy. For skin data, after about 1800 epochs both U-Net and mrU-Net started overfitting to the training data, and we stopped the training. Table 2 shows the number of iterations at the training stop point for each dataset and network separately. Fig. 3 shows the trends of validation sDSC during training for all four datasets. For each dataset, the validation accuracy of U-Net and mrU-Net were compared.

Table 2 The number of iterations for training the networks.

| Dataset | Network Model | Number of iterations (epochs) |
|---|---|---|
| Skin lesion | U-Net | 1700 |
| | mrU-Net | 1450 |
| LUNA | U-Net | 4950 |
| | mrU-Net | 4850 |
| DRIVE | U-Net | 4800 |
| | mrU-Net | 4950 |
| PROMISE12 | U-Net | 4750 |
| | mrU-Net | 4700 |

*3.2 Testing results*

To assess the segmentation performance of the proposed model, we compared the performance of the trained mrU-Net models with the corresponding trained U-Net models on the same test datasets. Table 3 shows the validation and testing results in terms of DSC, sensitivity rate, and specificity rate for the two networks. For each corresponding result pair, we applied a one-tailed *t*-test [31] with the null hypothesis that the modifications applied to the U-Net architecture did not improve the metric value. In Table 3, the metric values in bold show where the null hypothesis was rejected with α = 0.05. Figs 4, 5, 6, and 7 depict the qualitative segmentation results of U-Net and mrU-Net for the skin lesion, LUNA, DRIVE, and PROMISE12 datasets, respectively.

Table 3 Validation and testing performance of the U-Net and mrU-Net models in terms of DSC, sensitivity rate, and specificity rate on four different datasets. The metric values in bold indicate statistically significant differences detected between the U-Net and mrU-Net models (P < 0.05).

| Dataset | Network Model | Validation | | | Test | | |
|---|---|---|---|---|---|---|---|
| | | DSC | Sensitivity | Specificity | DSC | Sensitivity | Specificity |
| Skin lesion | U-Net | 66.3±24% | 81.6±26% | 96.1±5% | 70.4±20% | 88.6±19% | 89.1±12% |
| | mrU-Net | 66.7±24% | 81.0±26% | 96.3±4% | 70.6±20% | 87.7±18% | 89.5±11% |
| LUNA | U-Net | 96.4 ±2% | **98.9 ±0.4%** | **98.5±0.7%** | 97.3±2% | **98.8±0.6%** | **98.9±0.7%** |
| | mrU-Net | 97.6 ±1% | **99.3 ±0.4%** | **98.9±0.4%** | 97.9±2% | **99.2±0.5%** | **99.1±0.5%** |
| DRIVE | U-Net | **69.7±2%** | 93.2±2% | **93.4±0.9%** | 73.1±2% | 89.6±3% | 94.9±0.9% |
| | mrU-Net | **71.6±2%** | 93.5±1% | **94.0±1%** | 73.6±2% | 90.6±3% | 94.9±0.9% |
| PROMISE12 | U-Net | 82.9±10% | **93.5±9%** | 99.1±0.4% | 79.7±16% | **91.8±14%** | **98.9± 0.7%** |
| | mrU-Net | 82.2±10% | **88.3±15%** | 99.2±0.6% | 77.9±20% | **85.2±21%** | **99.1±0.7%** |

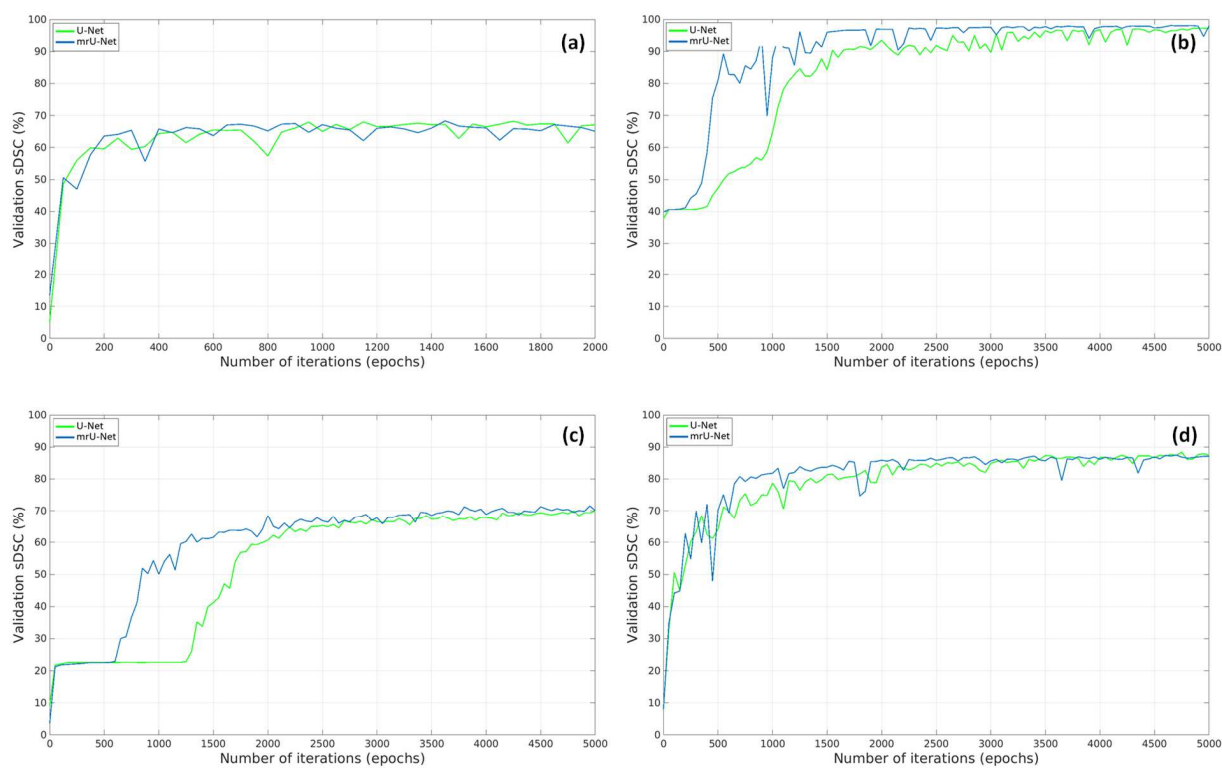

**Fig. 3** Validation sDSC for (a) skin, (b) LUNA, (c) DRIVE, and (d) PROMISE12 datasets during training.

## 4 Discussion

The proposed neural network architecture has provided a better or similar segmentation performance when compared to the U-Net architecture in the context of medical images. According to Fig. 3, the mrU-Net also showed a faster training rate on three of the four datasets used in this study, when compared to U-Net. These observations can be justified taking into account that the convolutional layers have been added to the U-Net architecture extracted features directly from the input image in multiple resolutions. Validation results presented in Table 3 and Fig. 3 show that after enough iterations both networks achieve the same level of accuracy. It means that U-Net can also extract the low-resolution features indirectly through the feature maps of higher levels but after more iterations. It implies the higher capability of mrU-Net in learning compared to U-Net.

The qualitative results have shown in Figs. 4 to 7 show that there are fewer false-positive objects detected by mrU-Net compared to U-Net. The slightly higher specificity rates of mrU-Net compared to U-Net in Table 3 also support this finding.

The low DSC values of the skin dataset could be explained by the high variability in the data. For the DRIVE dataset, due to the narrow and long structure of the retina vessels, DSC is not an appropriate error metric to evaluate the segmentation performance. It could be one reason for low DSC values observed for this dataset while the qualitative results in Fig. 6 show reasonable performances of the networks. In addition, as it is obvious from the figure, both U-Net and mrU-Net were unable to detect and segment small vessels. A DSC-based loss function could make it challenging for the networks to segment fine details like those small vessels. Another training optimization approach could improve the performance of the networks for such datasets.

*4.1 Limitations*

In this study, our goal was to compare the proposed architecture to U-Net architecture under the same training and testing conditions. Therefore, optimizations were not considered over the hyperparameter of the networks or post-processing of the output labels. The potential exists for the hypothesis to be tested and confirm obtained results after optimizing the hyperparameters of the networks through future work. A post-processing step could be helpful to improve the results. Furthermore, our observations are based on testing the proposed algorithm on four sample datasets. The hypothesis could be tested and verified more thoroughly with more datasets as a future research.

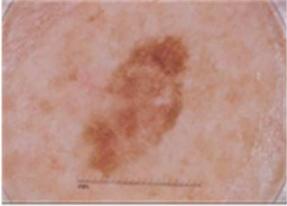

**Fig. 4** Qualitative segmentation results for skin lesion dataset on five sample images. Each row shows the results for one image. For each image, the ground truth segmentations are indicated in the second column and the U-Net and mrU-Net segmentations are exposed in the third and last columns, respectively. The DSC value for each algorithm result when compared to the ground truth is mentioned at the bottom of the segmentation label.

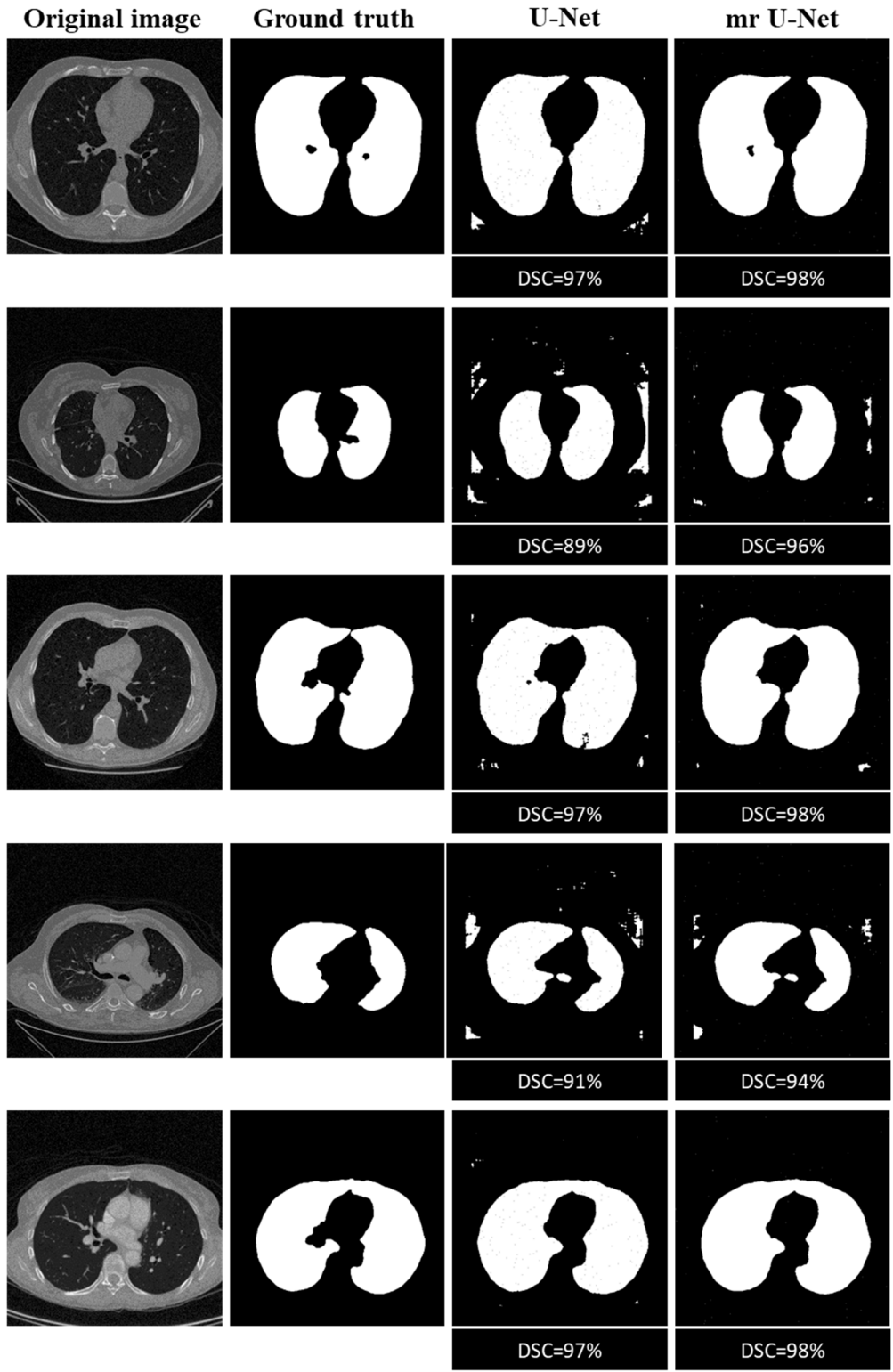

**Fig. 5** Qualitative segmentation results for the LUNA dataset on five sample images. Each row shows the results for one image. For each image, the ground truth segmentations are indicated in the second column and the U-Net and mrU-Net segmentations are exposed in the third and last columns, respectively. The DSC value for each algorithm result when compared to the ground truth is mentioned at the bottom of the segmentation label.

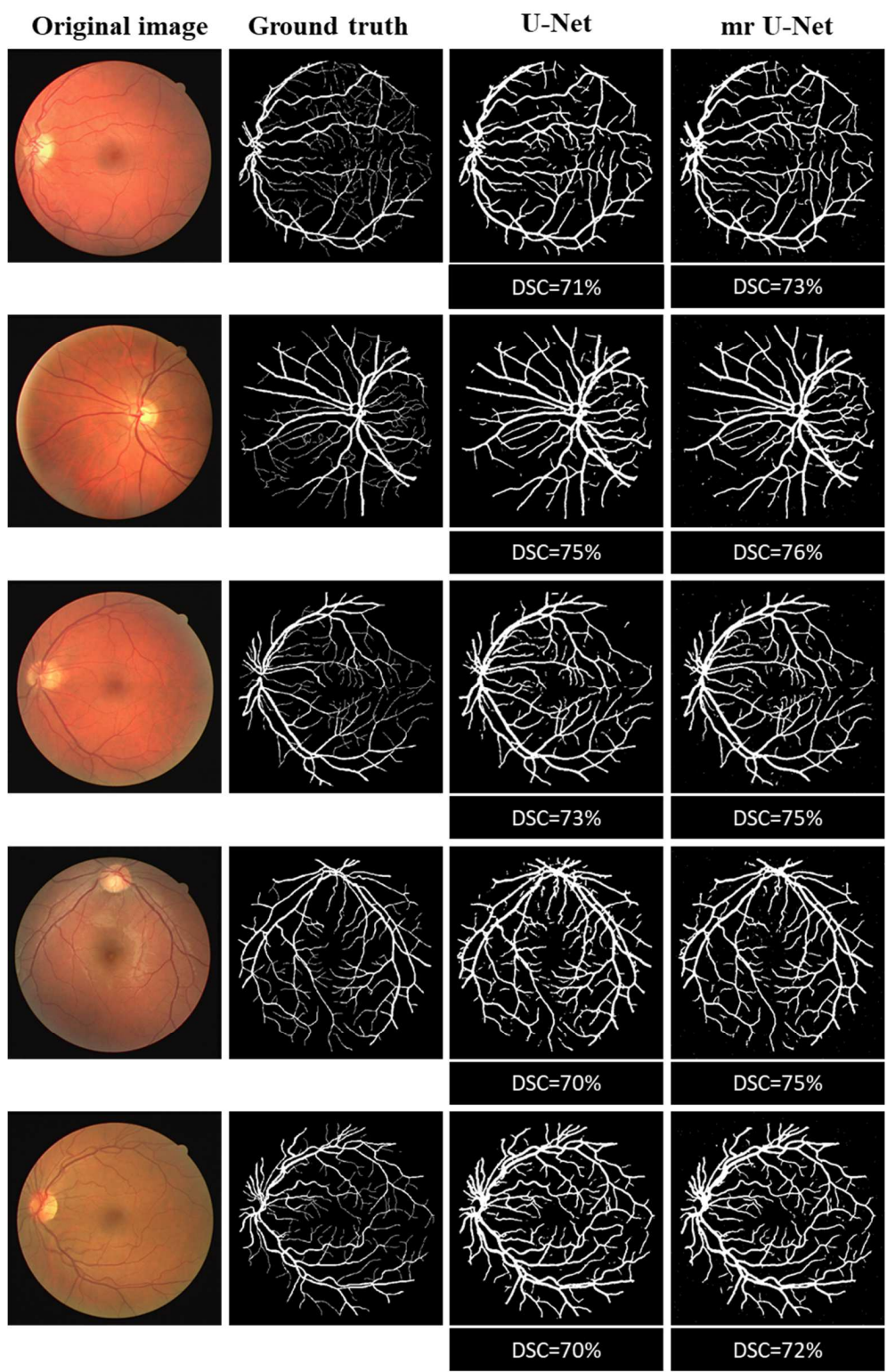

**Fig. 6** Qualitative segmentation results for retina blood vessel (DRIVE) dataset on five sample images. Each row shows the results for one image. For each image, the ground truth segmentations are shown in the second column and the U-Net and mrU-Net segmentations are shown in the third and last columns, respectively. The DSC value for each algorithm result when compared to the ground truth is mentioned at the bottom of the segmentation label.

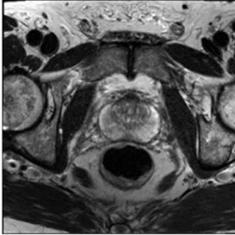

**Fig. 7** Qualitative segmentation results for the PROMISE12 dataset on five sample images. Each row shows the results for one image. For each image, the ground truth segmentations are indicated in the second column and the U-Net and mrU-Net segmentations are exposed in the third and last columns, respectively. The DSC value for each algorithm result when compared to the ground truth is mentioned at the bottom of the segmentation label.

*4.2 Conclusions*

We have proposed a new fully convolutional deep neural network architecture for the segmentation of medical images, which has been demonstrated high capability in learning and extracting image-derived features that could be helpful for image segmentation. We have tested our network on four different medical imaging datasets (skin lesion, LUNA, DRIVE, and PROMISE12) and have compared the performance of our algorithm against manual segmentation to evaluate the performance of the proposed method. We compared our network to the U-Net architecture as a state-of-the-art fully convolutional neural network widely used in medical image segmentation and demonstrated that it could be faster in learning and being trained in fewer iterations compared to the U-Net. Our proposed approach is expected to be deploy not only for medical image segmentation but also for object detection and segmentation on non-medical images.

*Disclosures*

The authors have no conflicts of interest to disclose.

*References*

1. W. L. Smith et al., "Prostate volume contouring: a 3D analysis of segmentation using 3DTRUS, CT, and MR," *International Journal of Radiation Oncology* Biology* Physics* **67**(4), 1238-1247 (2007).
2. M. R. Khokher, A. Ghafoor, and A. M. Siddiqui, "Image segmentation using multilevel graph cuts and graph development using fuzzy rule-based system," *IET image processing* **7**(3), 201-211 (2013).
3. R. Muthukrishnan, and M. Radha, "Edge detection techniques for image segmentation," *International Journal of Computer Science & Information Technology* **3**(6), 259 (2011).

4. V. K. Dehariya, S. K. Shrivastava, and R. Jain, "Clustering of image data set using k-means and fuzzy k-means algorithms," *2010 International Conference on Computational Intelligence and Communication Networks* 386-391 (2010).

5. S. N. Sulaiman, and N. A. M. Isa, "Adaptive fuzzy-K-means clustering algorithm for image segmentation," *IEEE Transactions on Consumer Electronics* **56**(4), 2661-2668 (2010).

6. D. Lin et al., "Scribblesup: Scribble-supervised convolutional networks for semantic segmentation," *Proceedings of the IEEE Conference on Computer Vision and Pattern Recognition* 3159-3167 (2016).

7. L. S. Davis, "A survey of edge detection techniques," *Computer graphics and image processing* **4**(3), 248-270 (1975).

8. A. C. Bovik, *Handbook of image and video processing*, Academic press (2010).

9. R. M. Haralick, and L. G. Shapiro, *Computer and robot vision*, Addison-wesley Reading (1992).

10. F. Milletari, N. Navab, and S.-A. Ahmadi, "V-net: Fully convolutional neural networks for volumetric medical image segmentation," *2016 Fourth International Conference on 3D Vision (3DV)* 565-571 (2016).

11. J. Long, E. Shelhamer, and T. Darrell, "Fully convolutional networks for semantic segmentation," *Proceedings of the IEEE conference on computer vision and pattern recognition* 3431-3440 (2015).

12. A. Prasoon et al., "Deep feature learning for knee cartilage segmentation using a triplanar convolutional neural network," *International conference on medical image computing and computer-assisted intervention* 246-253 (2013).

13. O. Ronneberger, P. Fischer, and T. Brox, "U-net: Convolutional networks for biomedical image segmentation," *International Conference on Medical image computing and computer-assisted intervention* 234-241 (2015).

14. X. Li et al., "H-DenseUNet: hybrid densely connected UNet for liver and tumor segmentation from CT volumes," *IEEE transactions on medical imaging* **37**(12), 2663-2674 (2018).

15. Y. Weng et al., "NAS-Unet: Neural Architecture Search for Medical Image Segmentation," *IEEE Access* **7**(44247-44257 (2019).

16. J. Dolz, C. Desrosiers, and I. B. Ayed, "IVD-Net: Intervertebral disc localization and segmentation in MRI with a multi-modal UNet," *International Workshop and Challenge on Computational Methods and Clinical Applications for Spine Imaging* 130-143 (2018).

17. Z. Guo et al., "Deep LOGISMOS: Deep learning graph-based 3D segmentation of pancreatic tumors on CT scans," *2018 IEEE 15th International Symposium on Biomedical Imaging (ISBI 2018)* 1230-1233 (2018).

18. L. Yu et al., "Automatic 3D cardiovascular MR segmentation with densely-connected volumetric convnets," *International Conference on Medical Image Computing and Computer-Assisted Intervention* 287-295 (2017).

19. H. Chen et al., "Voxresnet: Deep voxelwise residual networks for volumetric brain segmentation," *arXiv preprint arXiv:1608.05895* (2016).

20. K. Yan et al., "A propagation-DNN: Deep combination learning of multi-level features for MR prostate segmentation," *Computer methods and programs in biomedicine* **170**(11-21 (2019).

21. Z. Zeng et al., "RIC-Unet: An improved neural network based on Unet for nuclei segmentation in histology images," *Ieee Access* **7**(21420-21428 (2019).

22. Q. Jin et al., "RA-UNet: A hybrid deep attention-aware network to extract liver and tumor in CT scans," *arXiv preprint arXiv:1811.01328* (2018).

23. N. C. Codella et al., "Skin lesion analysis toward melanoma detection: A challenge at the 2017 international symposium on biomedical imaging (isbi), hosted by the international skin imaging collaboration (isic)," *2018 IEEE 15th International Symposium on Biomedical Imaging (ISBI 2018)* 168-172 (2018).

24. P. Tschandl, C. Rosendahl, and H. Kittler, "The HAM10000 dataset, a large collection of multi-source dermatoscopic images of common pigmented skin lesions," *Scientific data* **5**(180161 (2018).

25. "LUNA Dataset."
19

26. J. Staal et al., "Ridge-based vessel segmentation in color images of the retina," *IEEE transactions on medical imaging* **23**(4), 501-509 (2004).

27. G. Litjens et al., "Evaluation of prostate segmentation algorithms for MRI: the PROMISE12 challenge," *Medical image analysis* **18**(2), 359-373 (2014).

28. M. D. Zeiler, "ADADELTA: an adaptive learning rate method," *arXiv preprint arXiv:1212.5701* (2012).

29. L. R. Dice, "Measures of the amount of ecologic association between species," *Ecology* **26**(3), 297-302 (1945).

30. A. G. Lalkhen, and A. McCluskey, "Clinical tests: sensitivity and specificity," *Continuing Education in Anaesthesia Critical Care & Pain* **8**(6), 221-223 (2008).

31. R. F. Woolson, and W. R. Clarke, *Statistical methods for the analysis of biomedical data*, John Wiley & Sons (2011).

**Simindokht Jahangard** is a research assistant at the Isfahan University of Technology. She received her MS degrees in Robotics Engineering from the Amirkabir University of Technology of in 2014. She has worked on several image processing and machine vision projects such as object tracking, image retrieval in medicine images, methods for image-based plagiarism detection, vessel extraction and toad recognition.

**Maysam Shahedi** is a research associate at the University of Texas at Dallas. He has received his PhD in biomedical engineering from the University of Western Ontario, Canada. He also holds BSc and MSc degrees in electrical engineering from Isfahan University of Technology. His research interests are medical imaging, medical image processing, Image-guided Intervention, and machine learning.

**Caption List**

**Fig. 1** Sample images and their manual segmentation labels from the four datasets. Rows (a) and (c) show the images and rows (b) and (d) impart the ground truth of the corresponding images from the rows above them. Each column shows one dataset; form left to right, skin lesion, LUNA, DRIVE, and PROMISE12 dataset.

**Fig. 2** Four-level mrU-Net architecture. The hashed layers are added layers to the U-Net architecture. The numbers above the feature maps indicate the number of feature channels. The size of the original input image is 512×512 and it has either one channel for grayscale images or three channels for RGB images.

**Fig. 3** Validation sDSC for (a) skin, (b) LUNA, (c) DRIVE, and (d) PROMISE12 datasets during training.

**Fig. 4** Qualitative segmentation results for skin lesion dataset on five sample images. Each row shows the results for one image. For each image, the ground truth segmentations are indicated in the second column and the U-Net and mrU-Net segmentations are exposed in the third and last columns, respectively. The DSC value for each algorithm result when compared to the ground truth is mentioned at the bottom of the segmentation label.

**Fig. 5** Qualitative segmentation results for the LUNA dataset on five sample images. Each row shows the results for one image. For each image, the ground truth segmentations are indicated in the second column and the U-Net and mrU-Net segmentations are exposed in the third and last columns, respectively. The DSC value for each algorithm result when compared to the ground truth is mentioned at the bottom of the segmentation label.

**Fig. 6** Qualitative segmentation results for retina blood vessel (DRIVE) dataset on five sample images. Each row shows the results for one image. For each image, the ground truth segmentations are shown in the second column and the U-Net and mrU-Net segmentations are shown in the third and last columns, respectively. The DSC value for each algorithm result when compared to the ground truth is mentioned at the bottom of the segmentation label.

**Fig. 7** Qualitative segmentation results for the PROMISE12 dataset on five sample images. Each row shows the results for one image. For each image, the ground truth segmentations are indicated in the second column and the U-Net and mrU-Net segmentations are exposed in the third and last columns, respectively. The DSC value for each algorithm result when compared to the ground truth is mentioned at the bottom of the segmentation label.

**Table 1** The number of training, validation, and testing images for each dataset.

**Table 2** The number of iterations for training the networks.

**Table 3** Validation and testing performance of the U-Net and mrU-Net models in terms of DSC, sensitivity rate, and specificity rate on four different datasets. The metric values in bold indicate statistically significant differences detected between the U-Net and mrU-Net models ($P < 0.05$).